\documentclass[prd, twocolumn, nofootinbib, floatfix, preprintnumbers]{revtex4-1}

\usepackage[mathscr]{euscript}
\usepackage{amsmath}
\usepackage{graphicx}
\usepackage{dcolumn}
\usepackage{bm}
\usepackage{epsfig}
\usepackage{amssymb,latexsym,mathrsfs}
\usepackage{graphicx}
\usepackage{color}
\usepackage{hyperref}
\usepackage{float}
\usepackage{diagbox}
\usepackage{textpos}

\hypersetup{
    colorlinks=true,
    linkcolor=red,
    citecolor=blue,
} 

\newcommand{\be}{\begin{equation}}
\newcommand{\ee}{\end{equation}}
\newcommand{\bs}{\begin{split}} 
\newcommand{\bea}{\begin{eqnarray}}
\newcommand{\eea}{\end{eqnarray}}

\def\nn{\nonumber} 

\begin{document}

\title{1/4 BPS AdS$_3$/CFT$_2$}

\author{Yolanda Lozano$^1$,  Niall T. Macpherson$^{2,3}$, Carlos Nunez$^4$ and Anayeli Ramirez$^1$}
\affiliation{${}^1$ Department of Physics, University of Oviedo,
Avda. Federico Garcia Lorca s/n, 33007 Oviedo, Spain.}
\affiliation{${}^2$ SISSA International School for Advanced Studies,
Via Bonomea 265, 34136 Trieste and INFN sezione di Trieste,}
\affiliation{${}^3$ International Institute of Physics, Universidade Federal do Rio Grande do Norte,
Campus Universitario - Lagoa Nova, Natal, RN, 59078-970, Brazil,}
\affiliation{$^4$ Department of Physics, Swansea University, Swansea SA2 8PP, United Kingdom}

\begin{abstract}
We discuss new solutions in massive Type IIA supergravity with AdS$_3\times$S$^2$ factors, preserving ${\cal N}=(0,4)$ SUSY. We propose a duality with a precise family of  quivers that  flow to ${\cal N}=(0,4)$ fixed points at low energies. These quivers consist on two families of linear quivers coupled by matter fields. Physical observables such as the central charges provide stringent checks of the proposed duality. A formal mapping is presented connecting our backgrounds with those dual to six dimensional ${\cal N}=(1,0)$ CFTs, suggesting the existence of a flow across dimensions between the CFTs. 
\end{abstract}

\maketitle

\section{Introduction}
 An important  by-product of the Maldacena conjecture \cite{Maldacena:1997re}, has been a thorough study of supersymmetric and conformal field theories 
 (CFTs) in various dimensions. 
In particular, the last two decades  have witnessed  a large effort  in the classification of Type II or M-theory backgrounds with AdS$_{d+1}$ factors, see for example \cite{Gauntlett:2004zh,Gutowski:2014ova}. The  solutions are conjectured to be dual to CFTs in $d$ dimensions with different amounts of SUSY, that can then be studied holographically.

Major progress has been achieved when the CFT preserves half of the maximum number of allowed supersymmetries. For the case of ${\cal N}=2$ CFTs in four dimensions, the field theories studied in \cite{Witten:1997sc,Gaiotto:2009we} have holographic duals  discussed in \cite{Gaiotto:2009gz}, and further elaborated (among other works) in \cite{ReidEdwards:2010qs}-\cite{Bah:2019jts}. The case of five-dimensional CFTs was analysed from the field theoretical and holographic viewpoints in \cite{DHoker:2016ujz}-\cite{Bergman:2018hin}, among many other interesting works.   An infinite family of six-dimensional ${\cal N}=(1,0)$ CFTs was discussed from both the field theoretical and holographic points of view in \cite{Apruzzi:2015wna}-\cite{Hanany:1997gh}. For three-dimensional  ${\cal N}=4$ CFTs, the field theories presented in \cite{Gaiotto:2008ak} were discussed holographically in \cite{DHoker:2007hhe}-\cite{Lozano:2016wrs}, among other works.

The case of  two-dimensional CFTs  and their AdS duals  is particularly attractive, due to the interest that CFTs in two dimensions and  AdS$_3$ solutions present in other areas of theoretical physics. This applies in particular to the microscopical study of  black holes, where major progress has been achieved  \cite{Maldacena:1997de}-\cite{Couzens:2019wls}. This motivated various
 attempts at finding classifications of AdS$_3$  backgrounds and studying their dual CFTs   \cite{Witten:1997yu}-\cite{Deger:2019tem}.  ${\cal N}=(0,4)$ AdS$_3$ solutions remained however largely unexplored, with known cases following mostly from orbifoldings, string dualities or F-theory constructions. Two-dimensional CFTs with ${\cal N}=(0,4)$ supersymmetry constructed in the literature \cite{Haghighat:2013gba}-\cite{Hanany:2018hlz}
 await as well their holographic description. In this context, an important recent development has been the complete classification of AdS$_3$ solutions to massive IIA supergravity with small ${\cal N}=(0,4)$ supersymmetry (and SU(2) structure) achieved in \cite{Lozano:2019emq}. In this letter we  add a new entry to the dictionary between CFTs and string backgrounds with an AdS-factor by proposing explicit CFTs dual to these solutions. We define our CFTs as the IR fixed points of ${\cal N}=(0,4)$ UV finite two dimensional QFTs. These QFTs
 are
described by quivers, consisting of two long rows of gauge groups connected by hypermultiplets and Fermi multiplets. We show that the new background solutions to massive IIA supergravity constructed in \cite{Lozano:2019emq} contain the needed isometries to be dual to our CFTs. We give an example (further elaborated in \cite{LMNR2}, where additional examples can be found) that shows agreement between the field theory and holographic calculations of the central charge. Finally, we provide a formal mapping to the AdS$_7$ solutions constructed in \cite{Apruzzi:2013yva} that suggests the existence of a  flow across dimensions \cite{Donos:2014eua,Benini:2015bwz} 
 between the dual CFTs.
\section{The Geometry}
The backgrounds of massive Type IIA supergravity constructed in \cite{Lozano:2019emq} were proposed to be dual to ${\cal N}=(0,4)$ CFTs in two dimensions. These solutions have SL(2)$\times$SU(2) isometries and eight (four Poincar\'e plus four conformal) supercharges.
In this paper we will consider the particular case of the geometries in \cite{Lozano:2019emq} referred   therein as class I.  
In string frame they read,
\begin{eqnarray}
ds^2&=&g_1\!\bigg(ds^2(\!\text{AdS}_3\!)+g_2 ds^2(\text{S}^2)\!\bigg)+\!g_3 ds^2(\text{CY}_2)+\! \!\frac{ \!d\rho^2}{g_1},\nonumber \\
e^{-\Phi}&=& g_4,~ B_2=g_5 \text{vol}(\text{S}^2),\;\hat{F}_0=g_6,\;\;
\hat{F}_2=g_7\text{vol}(\text{S}^2),\nn\\
\hat{F}_4&=&g_8 d\rho \wedge\text{vol}(\text{AdS}_3)
+ g_9 \text{vol}(\text{CY}_2).\label{eq:background}
\end{eqnarray}
The functions $g_i$ are defined in terms of three functions, $u(\rho), \hat{h}_4(\rho), h_8(\rho)$, according to,
\begin{eqnarray}
& & g_1= \frac{u}{\sqrt{\hat{h}_4 h_8}},\;\;g_2= \frac{h_8\hat{h}_4 }{4 h_8 \hat{h}_4+(u')^2},\;\; g_3= \sqrt{\frac{\hat{h}_4}{h_8}},\nn\\
& & g_4=\frac{h_8^{\frac{3}{4}} }{2\hat{h}_4^{\frac{1}{4}}\sqrt{u}}\sqrt{4h_8 \hat{h}_4+(u')^2}, \;\;\; g_6= h_8', \;\; g_9=- \partial_{\rho}\hat{h}_4\nn
\end{eqnarray}
\begin{eqnarray}
& & g_5=\frac{1}{2}\left(-\rho+ 2\pi k+\frac{ u u'}{4 \hat{h}_4 h_8+ (u')^2} \right),\nn\\
& &  g_7=-\frac{1}{2}\bigg(h_8- h_8'(\rho-2\pi k)\bigg),\;\;g_8= \bigg(\partial_\rho\left(\frac{u u'}{2 \hat{h}_4}\right)+2 h_8\bigg).\nn
\end{eqnarray}
Notice that we have written the Page fluxes $\hat{F}= e^{-B_2} \wedge F$. We have also allowed for large gauge transformations 
 $B_2\to B_2 + {\pi k} \text{vol}(\text{S}^2)$, with $k=0,1,...., P$. The transformations are performed every time we cross a $\rho$-interval $[2\pi k, 2\pi (k+1)]$.  
The preservation of ${\cal N}=(0,4)$ supersymmetry implies $u''=0$. Away from localised sources Bianchi identities also impose $h_8''=0$ and ${\hat h}_4''=0$ \cite{Lozano:2019emq}.

Below, we present {\it new} solutions, defined piecewise in the intervals $[2\pi k ,2\pi (k+1)]$. For $\hat{h}_4$, $h_8$ we have,
 \begin{equation} \label{profileh4final}
\hat{h}_4(\rho)\!
                    =\!\Upsilon\!\! \left\{ \begin{array}{ccrcl}
                      \!\!\!\! \frac{\beta_0 }{2\pi}
                       \rho \!\!\!\!& \!\!\!\!0\leq \rho\leq 2\pi \\
                       \!\!\!               \alpha_k + \frac{\beta_k}{2\pi}(\rho-2\pi k), &\!\! 2\pi k\leq \rho \leq 2\pi(k+1)\\
                      \!\!\!\alpha_P-  \frac{\alpha_P}{2\pi}(\rho-2\pi P), & \!\!2\pi P\leq \rho \leq 2\pi(P+1),
                                             \end{array}
\right.
\end{equation} 
 \begin{equation} \label{profileh8final}
h_8(\rho)\!\!
                    =\!\!\left\{ \begin{array}{ccrcl}
                     \!\!\!\!  \frac{\nu_0 }{2\pi}
                       \rho & \!\!\!\!0\leq \rho\leq 2\pi \\
                                     \!\!\! \mu_k + \frac{\nu_k}{2\pi}(\rho-2\pi k) &\!\!\!2\pi k\leq \!\rho\! \leq 2\pi(k+1)\\
                    \!\!\!\!  \mu_P- \frac{\mu_P}{2\pi}(\rho-2\pi P) & 2\pi P\leq \rho \leq 2\pi(P+1),
                                             \end{array}
\right.
\end{equation}
while $u(\rho)=\frac{b_0}{2\pi}\rho$. 

\begin{figure}[h!]
    \centering

    {{\includegraphics[width=6cm]{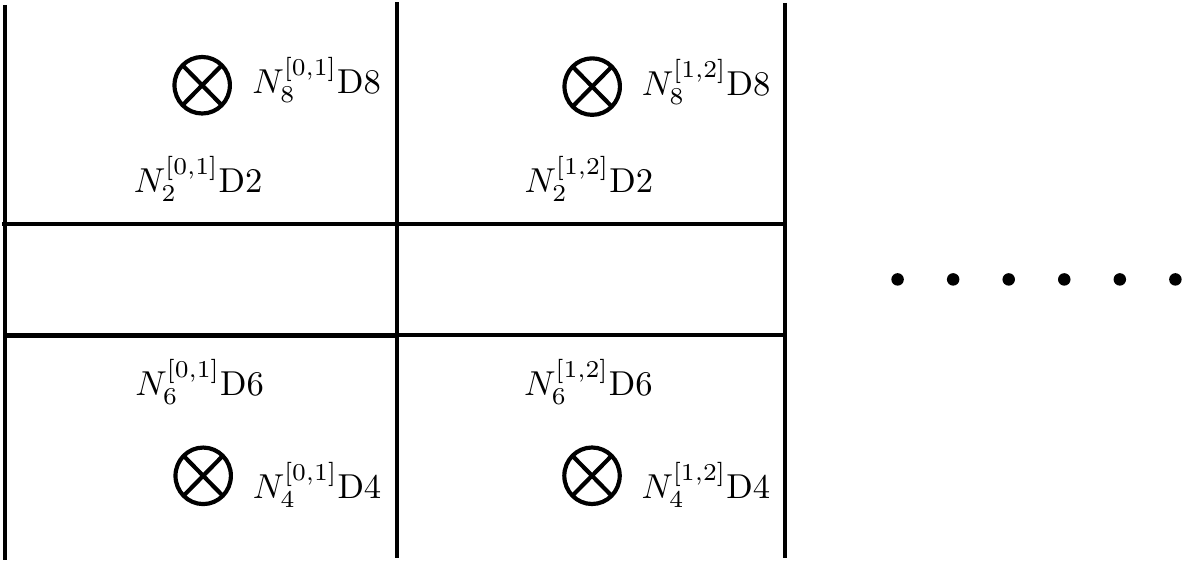} }}%

\caption{The generic Hanany-Witten set-up associated with our backgrounds. The vertical lines are NS-five branes. The horizontal lines represent D2 and wrapped D6 branes. The crosses indicate D4 and wrapped D8 branes.}
\label{xxy}

\end{figure}

Imposing the continuity of the Neveu-Schwarz (NS)-sector across the various intervals we find,
\begin{equation}
\mu_k=\sum_{j=0}^{k-1} \nu_j,\;\;\; \alpha_k=\sum_{j=0}^{k-1}\beta_j, \label{conticond}
\end{equation}
which also imply the continuity of  the functions $\hat{h}_4, h_8$ across intervals. The first derivatives present discontinuities at $\rho= 2\pi k$ where  D8 and D4 sources are located.
\\
{\it {\bf Page charges:}}
\\
The Page charges are important observable quantities characterising a supergravity solution. They are quantised, and are the ones that are related to the ranks of the gauge or global groups of the dual CFT. 
They are obtained integrating the Page fluxes, according to ${(2\pi)^{7-p} g_s \alpha'^{(7-p)/2}}Q_{Dp}= \int_{\Sigma_{8-p}} \hat{F}_{8-p}.$ \footnote{In what follows, we set $\alpha'=g_s=1$.}
This implies the 
quantisation of some of the constants in eqs.(\ref{profileh4final})-(\ref{profileh8final}). In the interval $[2\pi k, 2\pi(k+1)]$ we find,
\begin{eqnarray}
& & Q_{D8}=2\pi  F_0 =
  \nu_k,\;\;\; Q_{D6}=\frac{1}{2\pi}\int_{\text{S}^2} \hat{F}_2=
\mu_k.   \label{cargasxx}\\
& & Q_{D4}=\frac{1}{8\pi^3}\int_{\text{CY}_2} \hat{F}_4=\Upsilon \frac{\text{Vol(CY}_2)}{16\pi^4} \beta_k,\nonumber\\
& & Q_{D2}=\frac{1}{32\pi^5}\int_{\text{CY}_2\times \text{S}^2} \hat{F}_6=
 \Upsilon \frac{\text{Vol(CY}_2)}{16\pi^4} \alpha_k .\nonumber
\end{eqnarray}
We have used that the magnetic part $\hat{F}_{6,mag}=\hat{f}_6$ is 
\begin{equation}
\hat{f}_6=\frac{\Upsilon }{2}\left(h_4- h_4'(\rho-2\pi k)\right) \text{vol}(\text{S}^2)\wedge \text{vol}(\text{CY}_2) .\label{estax}
\end{equation} 
Besides, we count one NS-five brane every time we cross the value $\rho=2\pi k$ (for $k=1,....,P$). The total number of NS-five branes is $Q_{NS}=\frac{1}{4\pi^2}\int_{\rho\times \text{S}^2}H_3=P+1$.\\
The study of the Bianchi identities (see \cite{LMNR2} for the details), shows that dissolved in flux, we have  ``colour'' D2 and D6 branes. We also find that D4 and D8 branes play the role of ``flavour'', appearing explicitly as delta-function corrections of the Bianchi identities. For the interval $[2\pi(k-1), 2\pi k]$, we calculate 
\begin{eqnarray}
& & N_{D8}^{[k-1,k]}= \nu_{k-1}-\nu_k,\;\;\;\;\; N_{D4}^{[k-1,k]}= \beta_{k-1}-\beta_k,\label{numberofbranes}\\
& & 
N_{D6}^{[k-1,k]}= \mu_k=\sum_{i=0}^{k-1}\nu_i,\;\;\;\; N_{D2}^{[k-1,k]}= \alpha_k=\sum_{i=0}^{k-1}\beta_i.\label{numberofcolour}
\end{eqnarray}
We then have a Hanany-Witten brane set-up \cite{Hanany:1996ie}, that in the interval $[2\pi(k-1),2\pi k]$ (bounded by NS-five branes), has $N_{D6}^{[k-1,k]}, N_{D2}^{[k-1,k]}$ colour branes and  $N_{D8}^{[k-1,k]}, N_{D4}^{[k-1,k]}$ flavour branes (see table \ref{D6-NS5-D8-D2-D4-first} and  figure \ref{xxy}).
 \begin{table}[ht]
	\begin{center}
		\begin{tabular}{| l | c | c | c | c| c | c| c | c| c | c |}
			\hline		    
			& 0 & 1 & 2 & 3 & 4 & 5 & 6 & 7 & 8 & 9 \\ \hline
			D2 & x & x & &  &  &  & x  &   &   &   \\ \hline
			D4 & x & x &  &  &  &   &  & x & x & x  \\ \hline
			D6 & x & x & x & x & x & x & x  &   &   &   \\ \hline
			D8 & x & x &x  & x & x &  x &  & x & x & x  \\ \hline
			NS5 & x & x &x  & x & x & x  &   &   &  &  \\ \hline
		\end{tabular} 
	\end{center}
	\caption{$\frac18$-BPS brane intersection underlying our geometry. $(x^0,x^1)$ are the directions where the 2d CFT   (dual to our AdS$_3$) is defined. $(x^2, \dots, x^5)$ span the CY$_2$, on which the D6 and the D8-branes are wrapped. $x^6$ is the direction associated with $\rho$. Finally $(x^7,x^8,x^9)$ are the transverse directions realising the SU(2)-symmetry associated to S$^2$.}   
	\label{D6-NS5-D8-D2-D4-first}	
\end{table} 

To close the geometric part of our study, we use  the formalism in \cite{Macpherson:2014eza,Bea:2015fja} 
to calculate the holographic central charge. The result is (see \cite{LMNR2} for a derivation),
\begin{equation}
c_{hol}=\frac{3\pi}{2G_N}\text{Vol}(\text{CY}_2) \int_0^{2\pi(P+1)} \hat{h}_4 h_8 d\rho. \label{centralhol}
\end{equation}
Since the backgrounds have localised singularities, associated with the presence of D-branes, observables calculated using the geometry are trustable as long as the numbers $\nu_k, \beta_k,  b_0, P$ are large.
\section{The Field Theory}
\begin{figure}[h!]
    \centering
    {{\includegraphics[width=3cm]{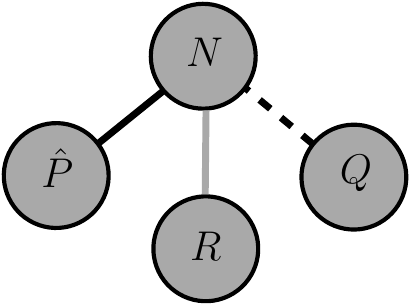} }}%

\caption{The building block of our theories. 
}
\label{explanation}

\end{figure}

\begin{figure}[h!]
    \centering
    {{\includegraphics[width=7cm]{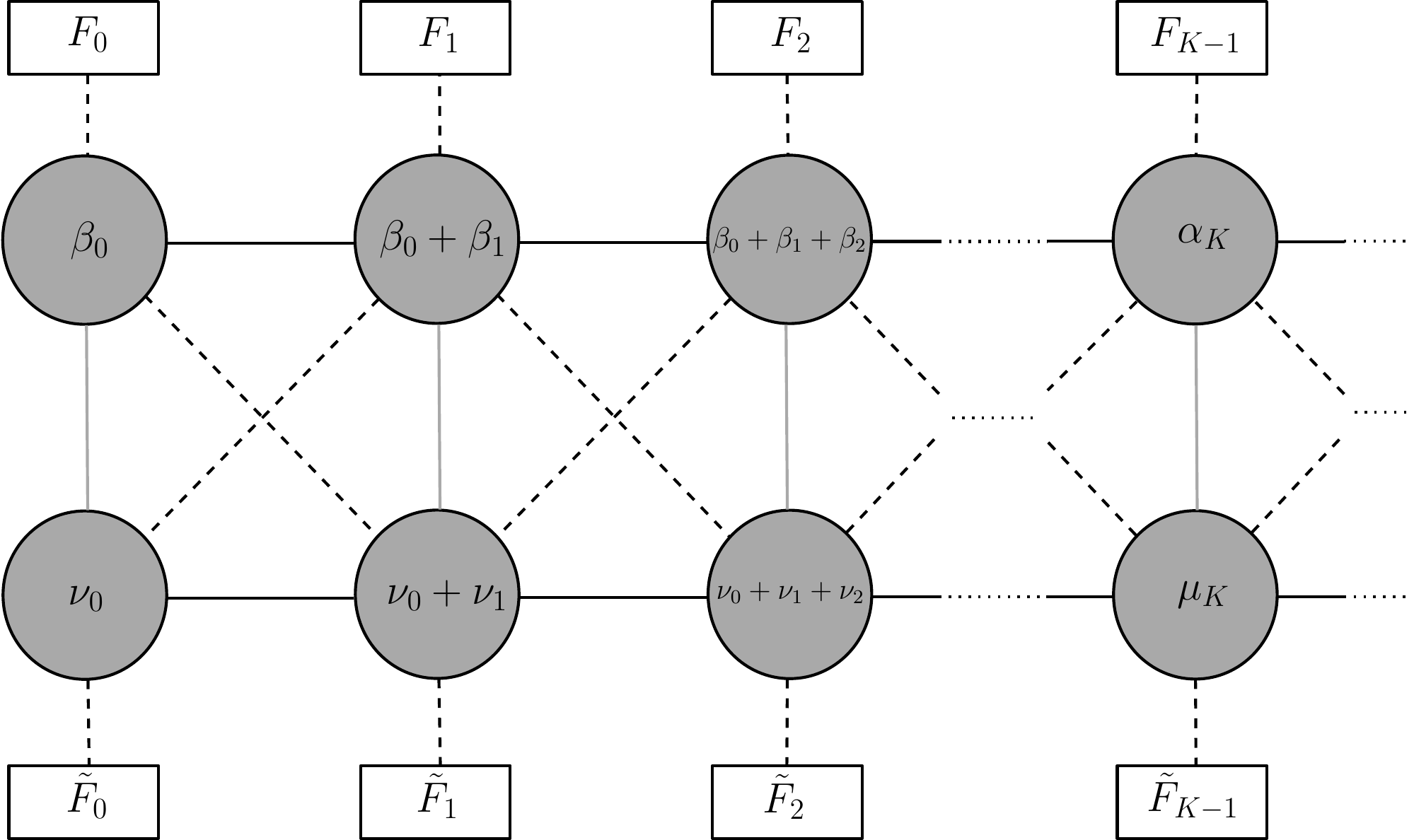} }}%

\caption{ A generic quiver field theory whose IR is dual to the holographic background defined by eqs.\eqref{eq:background}-\eqref{profileh8final}. 
}
\label{figurageneral}

\end{figure}

\begin{figure}[h!]
    \centering
    {{\includegraphics[width=9cm]{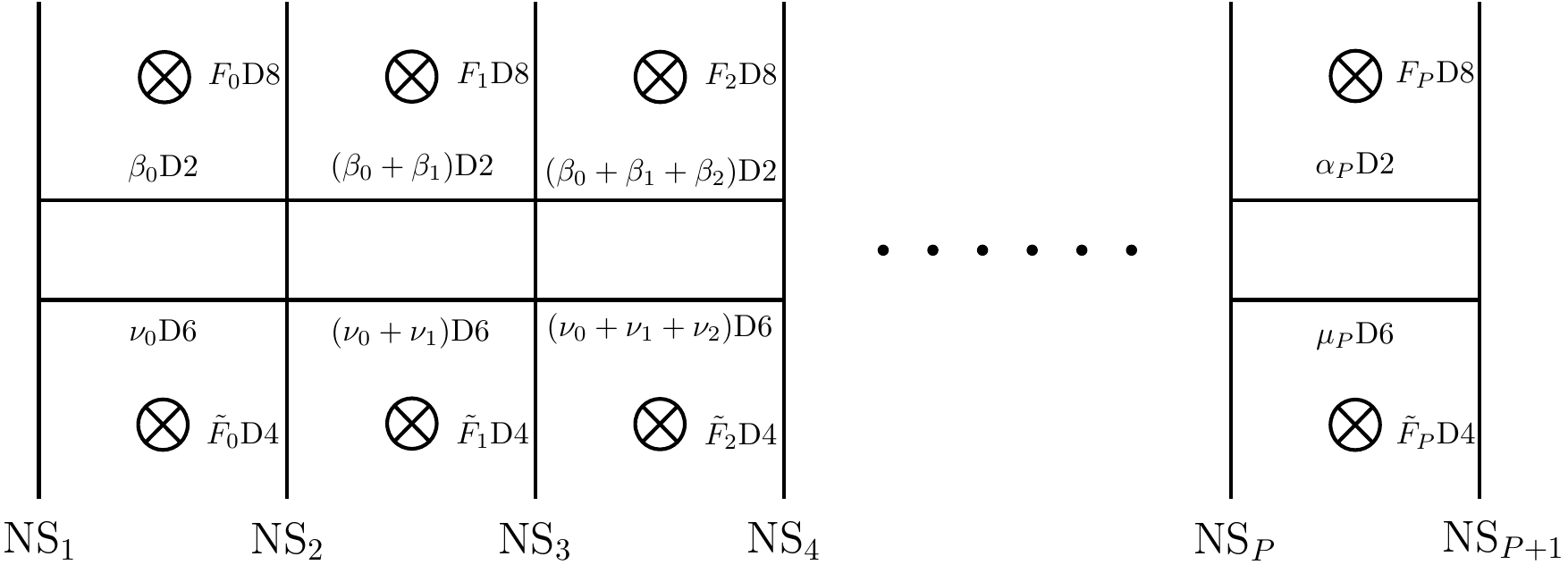} }}%

\caption{Hanany-Witten set-up associated with our generic quiver in figure \ref{figurageneral}. 
}
\label{vvvbb}

\end{figure}

In this section we discuss the two-dimensional CFTs dual to the backgrounds  given by eqs.\eqref{eq:background}-\eqref{profileh8final}.  They are defined as the strongly coupled IR fixed points of  QFTs that in the  (weakly coupled) UV are constructed from the ``building block'' depicted in 
figure \ref{explanation}.
We have an SU($N$) gauge group with the matter content of a two-dimensional  ${\cal N}=(4,4)$ vector multiplet in the adjoint of SU($N$). This gauge group is joined with other (gauge or global) symmetry groups SU($\hat{P}$), SU($R$) and SU($Q$). The connection with the SU($\hat{P}$) symmetry group is mediated by  ${\cal N}=(4,4)$ hypermultiplets running over the black solid line, that with the SU($R$) symmetry group via   ${\cal N}=(0,4)$ hypermultiplets that propagate over the grey lines, and that with SU($Q$) via ${\cal N}=(0,2)$ Fermi multiplets that run over the dashed line\footnote{Notice that ${\cal N}=(0,2)$ Fermi multiplets are allowed in ${\cal N}=(0,4)$ theories as long as they do not transform under the $SO(4)$ R-symmetry (see \cite{Witten:1994tz}).}. All these multiplets transform in the bifundamental representation of the gauge groups. A similar (but not the same) field content was used in \cite{Gadde:2015tra}.
\\
The cancellation of gauge anomalies constrains the ranks of the different symmetry groups. Using the contribution to the gauge anomaly
coming from each multiplet, see \cite{Franco:2015tna}, we find that for the SU($N$) gauge group the cancellation of the anomaly imposes,
\begin{equation}
2R=Q \, .\label{noanomaly}
\end{equation}
 Our quiver gauge theories are then obtained by ``assembling'' the building blocks of figure \ref{explanation} such that there is anomaly cancellation for all gauge groups.
\\
In turn, the central charge of the IR CFT is calculated by associating it with the correlation function of U(1)-R-symmetry currents (computed in the UV-description above).
At the conformal point, the (right-moving) central charge is  related to the, two point, U(1)$_R$ current correlation function, such that (see \cite{Putrov:2015jpa}), 
\begin{equation}
c= 6 (n_{hyp}- n_{vec}).\label{centralfinal}
\end{equation}
The central charge is then obtained by counting the number of ${\cal N}=(0,4)$ hypermultiplets and substracting the number of ${\cal N}=(0,4)$ vector multiplets in the UV description. Note that the $SU(2)$ R-symmetry does not mix with the Abelian flavour symmetries, and it is not necessary to go through a c-extremisation procedure  \cite{Benini:2013cda}. 
\\
{\bf The proposed duality:}
\\
Our proposal relates the backgrounds in eqs.\eqref{eq:background}-\eqref{profileh8final}
with quiver field theories obtained by assembling the building block depicted in figure \ref{explanation}. For  generic functions $\hat{h}_4, h_8$ this results in the quiver shown in figure \ref{figurageneral}, associated to the Hanany-Witten set-up in figure \ref{vvvbb}.
The reader can check that the cancellation of gauge anomalies implies
for a generic $\text{SU}(\alpha_k)$ colour group, in the interval $[2\pi(k-1), 2\pi k]$
\begin{equation}
F_{k-1}+\mu_{k+1}+\mu_{k-1}=2\mu_k\to  F_{k-1}=\nu_{k-1}-\nu_k,
\end{equation}
which, according to \eqref{numberofbranes},  is precisely the number of flavour D8 branes in the $[2\pi(k-1), 2\pi k]$ interval of the brane set-up. Things work analogously if D6 are replaced by D2  (or $\mu_k \leftrightarrow \alpha_k$) and D8 by D4 ($\nu_k \leftrightarrow \beta_k$), and we work with a generic lower-row gauge group $\text{SU}(\mu_k)$.
%

\begin{figure}[h!]
    \centering
    {{\includegraphics[width=6cm]{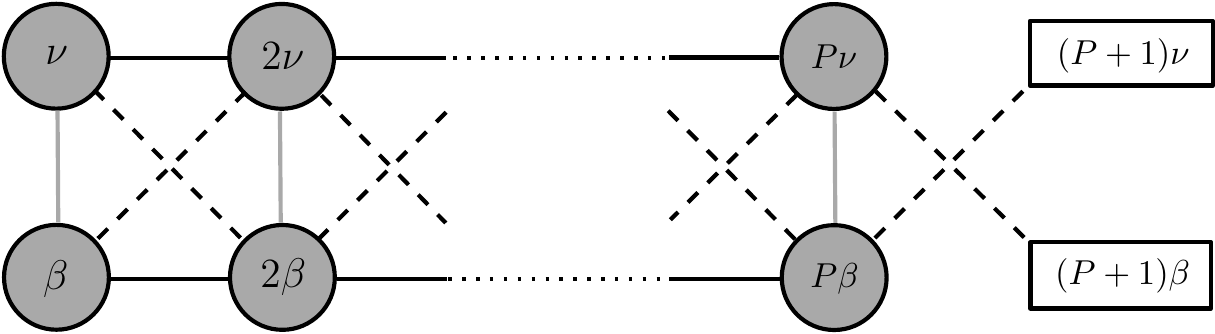} }}%

\caption{Quiver consisting of two rows of linearly increasing colour groups, terminated with the addition of flavour groups.
}
\label{example}

\end{figure}

 %
 The central charge of the quiver is calculated using expresion (\ref{centralfinal}). We find,
\begin{eqnarray}
& & \frac{c}{6}\!\!= \!\!  \sum_{j=1}^P\! \Bigl(\!\alpha_j\mu_j -\!\alpha_j^2-\!\mu_j^2\!+2\!\Bigr) \!\!+\!\!\!\sum_{j=1}^{P-1} \!\!\Bigl(\!\alpha_j \alpha_{j+1}\!+\!\mu_j\mu_{j+1}\!\Bigr)\!. \label{batigol}
\end{eqnarray}
In \cite{LMNR2}  we present various examples in which this expression agrees with the holographic central charge computed according to \eqref{centralhol}. This should hold in the limit in which the 
number of nodes $P$ and the ranks of each gauge group $\alpha_i,\mu_i$ are large, which is when the supergravity backgrounds are trustable. We present one such example below.
Notice that both the global symmetries and isometries (space-time, SUSY and flavour), as well as the ranks of the gauge (colour) groups do match in both descriptions, the latter being given by the numbers $\alpha_k,\mu_k$ in \eqref{conticond}. 
\\
{\bf An example:}
\\
Let us discuss an example that illustrates the duality proposed above. We consider the quiver with two rows of linearly increasing colour groups depicted in Figure \ref{example}.
For an intermediate gauge node SU($k \nu$)  we have $ Q= 2 k\beta , R=k\beta$. This implies that \eqref{noanomaly} is satisfied and any generic intermediate gauge group is not anomalous. If we refer to the last gauge group in the upper-row  SU($P\nu$) we have that $Q=(P+1)\beta+(P-1)\beta=2P\beta$ and $R=P\beta$. As a consequence  \eqref{noanomaly} is satisfied and the gauge group SU($P\nu$)  is also not anomalous. The same occurs for the lower-row gauge groups.

The counting of $(0,4)$ hypermultiplets and vector multiplets gives
\begin{eqnarray}
& &  n_{vec}=\sum_{j=1}^P \Bigl(j^2(\nu^2+\beta^2)-2\Bigr),\nn\\
& &
 n_{hyp}=\sum_{j=1}^{P-1} j(j+1)(\nu^2+\beta^2)+ \sum_{j=1}^{P} j^2\nu\beta,
\end{eqnarray}
from which the central charge of the IR CFT can be computed,
\begin{eqnarray}
c&=&6\nu\beta (\frac{P^3}{3} +\frac{P^2}{2} +\frac{P}{6}) - 3(\nu^2+\beta^2)(P^2+P)+12 P\nonumber\\
&\sim& 2\nu\beta P^3.\label{centralchargeexampleII}
\end{eqnarray}
In turn, the holographic description of the system is in terms of the functions,
\begin{equation} \label{profileh8exampleII}
h_8(\rho)
                    =\left\{ \begin{array}{ccrcl}
                       \frac{\nu }{2\pi}
                       \rho & 0\leq \rho\leq 2 \pi P\\
                      \frac{\nu P}{2\pi}(2\pi( P+1) -\rho) & 2\pi P\leq \rho \leq 2\pi(P+1), \nonumber
                                             \end{array}
\right.
\end{equation}
  \begin{equation} \label{profileh4exampleII}
\hat{h}_4(\rho)\!
                    =\!\Upsilon\!\!\left\{ \begin{array}{ccrcl}
                       \frac{\beta }{2\pi}
                       \rho & 0\leq \rho\leq 2\pi P\\
                      \frac{\beta P}{2\pi}(2\pi (P+1) -\rho) & 2\pi P\leq \rho \leq 2\pi(P+1).\nonumber
                                             \end{array}
\right.
\end{equation}
Using \eqref{centralhol} and a convenient choice for the constant $\Upsilon$,  gives rise to the holographic central charge,
\begin{equation}
c_{hol}=2\nu\beta P^3(1+\frac{1}{P})\sim 2 \nu\beta P^3.\label{ccholii}
\end{equation}
We thus find perfect agreement between the field theory and holographic calculations. In \cite{LMNR2} other examples of dual holographic pairs are discussed that provide stringent support to our proposed duality.
\\
\section{Mapping to $\text{AdS}_7$ backgrounds}
A sub-class of the solutions discussed in \cite{Lozano:2019emq} can be related to the AdS$_7$ solutions to massive IIA supergravity constructed in \cite{Apruzzi:2013yva}. As opposed to the mappings in \cite{Apruzzi:2015wna}, this mapping is not one-to-one, due to the presence of additional D2-D4 branes in the AdS$_3$ solutions, whose backreaction introduces extra 4-form and 6-form fluxes, and reduces the supersymmetries by a half. Using this map it is possible to give an interpretation to the 2d CFTs dual to the AdS$_3$ solutions as associated to D2-D4 defects in the D6-NS5-D8 brane set-ups dual to the AdS$_7$ solutions in \cite{Apruzzi:2013yva}, wrapped on the CY$_2$. Thus, the word defect is here used to indicate the presence of extra branes in Hanany-Witten brane set-ups that would otherwise arise from compactifying higher dimensional branes. 

 The explicit map, discussed in detail in \cite{LMNR3}, reads, 
  \begin{eqnarray}
  \label{fieldthdir}
& &   \rho\leftrightarrow2\pi z,
~    u\leftrightarrow\alpha,
~    h_8\leftrightarrow-\frac{\ddot{\alpha}}{81 \pi^2}, 
  ~  \hat{h}_4\leftrightarrow \frac{81}{8} \alpha,\nn\\
  %
    & & ds^2(\text{AdS}_3)+\frac{3^4}{2^3}ds^2(\text{CY}_2)\leftrightarrow \,4\, ds^2(\text{AdS}_7)\;\;\;.
\end{eqnarray}
This transforms the original backgrounds in \eqref{eq:background} into the AdS$_7$ backgrounds constructed in \cite{Apruzzi:2013yva} 
    %
\begin{eqnarray}
\label{metricAdS7alpha}
\!\!\frac{ds_{10}^2}{\pi\sqrt{2} }&=&8\sqrt{-\frac{\alpha}{\ddot{\alpha}}} ds^2(\text{AdS}_7)\!+\!\sqrt{-\frac{\ddot{\alpha}}{\alpha}}dz^2\!+\!\frac{\alpha^{3/2}(-\ddot{\alpha})^{1/2}}{\Delta}ds^2(\text{S}^2),\nn \\
e^{2\Phi}&=&2^{5/2}\pi^5 3^8\frac{(-\alpha/\ddot{\alpha})^{3/2}}{\dot{\alpha}^2-2\alpha\ddot{\alpha}},~~ F_0=-\frac{\dddot{\alpha}}{162\pi^3}\nn\\[1mm]
B_2&=&\pi \Bigl(-z+k+\frac{\alpha\dot{\alpha}}{\Delta}\Bigr)\text{vol}(\text{S}^2),\!\!\\
\hat{F}_2&=&\frac{1}{162\pi^2}\Bigl(\ddot{\alpha} -\dddot{\alpha}(z-k)
\Bigr)\text{vol}(\text{S}^2),~ \Delta=\dot{\alpha}^2-2\alpha\ddot{\alpha},\nn\label{F2AdS7alpha}
\end{eqnarray}
where the function $\alpha(z)$ satisfies the equation 
$
\dddot{\alpha}=-162\pi^3 F_0.
$
As analysed in \cite{Apruzzi:2013yva},  $\alpha(z)$ encodes the information about the 6d (1,0) dual CFT, which is realised in a D6-NS5-D8 Hanany-Witten set-up.
 Using the mapping defined by (\ref{fieldthdir}) it is possible to obtain an AdS$_7$ solution in the class of \cite{Apruzzi:2013yva} from an AdS$_3$ solution. In turn, the D6-NS5-D8 {\it sector} of the AdS$_3$ solution is obtained by compactifying  the D6-NS5-D8 branes that underlie the AdS$_7$ solution on the CY$_2$, while it is necessary to add the D2-D4 {\it sector}, encoded by the functions $u$ and $h_4$ (see \cite{LMNR3} for the details) to achieve conformality and fully determine the AdS$_3$ solution.
 
The holographic central charge of the 6d CFTs dual to the AdS$_7$ solutions was computed in \cite{Nunez:2018ags},
  \begin{equation}
 c_{\text{AdS}_7}=\frac{1}{G_N}\frac{2^4}{3^8}\int dz (-\alpha \ddot{\alpha}).
 \end{equation}
 In turn, the holographic central charge of the 2d CFTs is given in \eqref{centralhol}. Using the mapping given by (\ref{fieldthdir}) this becomes,
 \begin{equation}
 \label{ces}
 c_{\text{AdS}_3}\leftrightarrow\frac{3}{2^3 G_N}{\rm \text{Vol}}(\text{CY}_2)\int dz (-\alpha \ddot{\alpha})=\frac{3^9}{2^7}
 {\rm Vol}(\text{CY}_2) \,c_{\text{AdS}_7}\, .\nn
 \end{equation}
 This kind of relation is ubiquitous when calculating the holographic central charges for ``flows across dimensions''.
 Our result strongly suggests that we can obtain our CFTs by compactifying  the D6-NS5-D8 system underlying the 6d (1,0) CFT on a  $\text{CY}_2$. Conformality in the lower dimensional theory however requires the presence of ``defect'' D2 and D4 branes, represented by the fluxes $F_4, F_6$ in \eqref{eq:background}. Flows of this type were studied in
 \cite{Dibitetto:2017tve},\cite{Dibitetto:2017klx}, but these do not reach an AdS$_3$ fixed point. It would be interesting to find the explicit RG flows that deform the six-dimensional ${\cal N}=(1,0)$ CFT to reach a two-dimensional ${\cal N}=(0,4)$ conformal fixed point in the IR.
\section{Conclusions}
This letter presents a new entry in the mapping between CFTs and AdS-supergravity backgrounds, for the case of  two-dimensional (small) ${\cal N}=(0,4)$  CFTs and backgrounds with AdS$_3\times$S$^2$ factors. We have reported new solutions of the type AdS$_3\times$S$^2\times$CY$_2$, belonging to class I in the classification in  \cite{Lozano:2019emq}, with compact CY$_2$ and  piecewise continuous defining functions. We have proposed explicit 2-d dual CFTs based on the Hanany-Witten set-ups implied by the Page charges of the solutions. We matched the background isometries and the global symmetries (both space-time and flavour) of the CFTs, and checked the agreement between the holographic and field theory central charges. The CFTs are defined as the IR limit of UV finite long  quivers with $(0,4)$ SUSY, that generalise 2-d (0,4) quivers previously discussed in the literature  \cite{Gadde:2015tra,Hanany:2018hlz}.  We presented a map between a sub-class of the solutions in  \cite{Lozano:2019emq} and the AdS$_7$ backgrounds dual to six dimensional ${\cal N}=(1,0)$ CFTs \cite{Apruzzi:2013yva}-\cite{Cremonesi:2015bld}. This mapping suggests the possibility of finding an RG flow across dimensions between the dual  CFTs.\\
This paper just scratches the surface of a rich line of work.
 In the forthcoming papers \cite{LMNR2,LMNR3} we will present various checks of our proposed duality. As a by-product we will obtain explicit  completions of the background obtained via non-Abelian T-duality on AdS$_3\times$S$^3\times$CY$_2$, along the lines of \cite{Lozano:2016kum,Lozano:2016wrs,Lozano:2017ole,Itsios:2017cew,Lozano:2018pcp}.

\section*{Acknowledgements} 
We would like to thank Giuseppe Dibitetto,  Gaston Giribet, S. Prem Kumar, Daniel Thompson, Alessandro Tomasiello and Stefan Vandoren for very useful discussions.
YL and AR are partially supported by the Spanish government grant PGC2018-096894-B-100 and by the Principado de Asturias grant FC-GRUPIN-IDI/2018/000174. NTM is funded by the Italian Ministry of Education, Universities and Research under the Prin project ``Non Perturbative Aspects of Gauge Theories and Strings'' (2015MP2CX4) and INFN. CN is Wolfson Fellow of the Royal Society. AR is supported by CONACyT-Mexico. We would like to acknowledge the Mainz Institute for Theoretical Physics (MITP) of the DFG Cluster of Excellence PRISMA$^{+}$ (Project ID 39083149) and the Theory Unit at CERN for their hospitality and partial support during the development of this work. \\
~\\
Email: ylozano@uniovi.es, ntmacpher@gmail.com, c.nunez@swansea.ac.uk, anayelam@gmail.com

\end{document}